%% file: edu_note_els1.tex
\documentclass[twocolumn,5p, authoryear, round]{elsarticle}
    \makeatletter
    \def\ps@pprintTitle{%
       \let\@oddhead\@empty
       \let\@evenhead\@empty
       \def\@oddfoot{\reset@font\hfil\thepage\hfil}
       \let\@evenfoot\@oddfoot
    }
    \makeatother
\usepackage{flushend}
\usepackage{amssymb}
\usepackage{float}
\usepackage{amsmath}
\usepackage{enumerate}
\usepackage{bm}
\usepackage{adjustbox}
\usepackage{booktabs}
\usepackage{dcolumn} 
\newcolumntype{d}[1]{D{.}{.}{#1}}
\usepackage[flushleft]{threeparttable} 

\usepackage[table,xcdraw]{xcolor}

\usepackage[labelfont=bf, font=footnotesize, labelsep=period]{caption}
\usepackage{subcaption, multicol}

\graphicspath{{"C:/Users/Fabio M/Documents/Research/UNSW Res/Marriage/Graphs/"}}

\usepackage[colorlinks=true,allcolors=rgb 15 128 172, pagebackref=true]{hyperref} 
\renewcommand*{\backref}[1]{}
\renewcommand*{\backrefalt}[4]{%
    \ifcase #1 (Not cited.)%
    \or        [Cited on page~#2.]%
    \else      [Cited on pages~#2.]%
    \fi}
\usepackage{url}
	
\usepackage[utf8]{inputenc}

\usepackage{scalerel,stackengine}
\stackMath
\newcommand\reallywidehat[1]{%
\savestack{\tmpbox}{\stretchto{%
  \scaleto{%
    \scalerel*[\widthof{\ensuremath{#1}}]{\kern.1pt\mathchar"0362\kern.1pt}%
    {\rule{0ex}{\textheight}}
  }{\textheight}%
}{2.4ex}}%
\stackon[-6.9pt]{#1}{\tmpbox}%
}
\usepackage{wrapfig}  
\usepackage{verbatim}  
\usepackage{lscape}  

\begin{document}

\begin{frontmatter}

\title{Increasing the price of a university degree does not significantly affect enrolment if income contingent loans are available: evidence from HECS in Australia}






\author[First,Second]{Fabio I. Martinenghi}
\ead{f.martinenghi@unsw.edu.au}
\ead[url]{https://fabitmart.github.io/}
\address[First]{School of Economics, University of New South Wales, Australia.}
\address[Second]{Australian Research Council Centre of Excellence for Children and Families over the Life Course, Australia.}

\begin{abstract}
I provide evidence that, when income-contingent loans are available, student enrolment in university courses is not significantly affected by large increases in the price of those courses. I use publicly available domestic enrolment data from Australia. I study whether large increases in the price of higher education for selected disciplines in Australia in 2009 and in 2012 was associated with changes in their enrolment growth. I find that large increases in the price of a course did not lead to significant changes in their enrolment growth for that course.\\

\end{abstract}


\end{frontmatter}
\section{Introduction}
In Australia, the federal government provides financial aid for its undergraduate domestic students (hereafter referred to as ``students'') in several ways. A complete exposition of how this system works can be found in \cite{CHAPMAN06} and a summary in \cite{CHAP_LEIGH09}. I will only outline the essential features of this system. Under the \textit{Higher Education Support Act 2003} \citep{HE03}, the government pays their degrees upfront and in full. Part of this government payment is a subsidy, while the remainder is a loan. This latter part is called ``student contribution'' and it is the total amount that students will pay for their university degree. The student contribution is repaid via the Higher Education Contribution Scheme (HECS). Under this scheme, students will gradually repay the government only after they earn a sufficiently high income. In short, the government gives prospective students an income-contingent loan\footnote{On income-contingent loans, see for instance \citep{CHAP_STIG_14} and \citep{BRITTON2019}.}. Moreover, HECS is progressive so that graduates with higher incomes will pay the loan back at higher rates.

The federal government sets the \textit{maximum student contribution} that Australian universities can charge for each discipline, or, equivalently, the highest amount that students will have to pay for studying a certain discipline.
Each discipline is assigned to one of three bands, which are associated with different maximum student contributions. The \textit{Higher Education Support Act 2003} \citep{HE03} introduced a fourth band, the ``National Priorities''. This band is assigned to those disciplines that the Commonwealth government is seeking to promote, including by reducing their maximum student contribution. Between 2005-2009, \textit{Nursing} and \textit{Education} were declared national priorities. Between 2009-2012, all the Natural and Physical Sciences (which include Mathematical Sciences) were declared national priorities. 

After 2009, when Nursing and Education were no longer considered a national priority, their maximum student contributions increased by approximately 25\%, from \$4249 to \$5310. After 2012, when the Natural and Physical Sciences were no longer considered a national priority, their maximum student contributions increased by approximately 85\%, from \$4520 to \$8363.

The HECS system was designed to make higher education financially accessible to every Australian. The income-contingent nature of the loan is a feature aimed at helping students with the cost of their degrees. It is possible that the HECS system has relieved students from funding concerns to the point that they have become insensitive to changes in their student contribution. If this were true, it would imply that under a system like HECS, where income-contingent loans are available to students, a government cannot effectively use university fees (maximum student contribution) as a policy instrument to boost demand for certain disciplines. In other words, changing the price of a degree to encourage or discourage enrolment in it would not work. 

In this note, I test whether the event of removing a discipline from the National Priority band \--- hence increasing its maximum student contribution \--- is associated which a change in the student enrolment for that discipline. The change in student enrolment is measured as a change in the slope of the student enrolment trend after the event.
I find that moving a discipline from the National Priority band had no significant impact on student enrolment. 

\section{Methods}
\subsection{Data}
I collect and combine publicly available data on commencing domestic undergraduate students and on the Annual Course Contribution Value (ACCVAL) tables between 2004-2018. The ACCVAL tables provide information on maximum student contribution by discipline. Years 2004 and 2005 have been dropped from the analysis due to inconsistencies in how cross-institutional undergraduate students were categorised. This categorisation differs from the rest of the sample and creates uncertainty around the yearly number of commencing undergraduate students. The dataset and the R code to reproduce it\footnote{This includes an automated download of all the data and their manipulation.} are available on my website \href{https://fabitmart.github.io/portfolio/3-au-edu-wp/}{(link)}.

\subsection{Empirical framework}
While commencing students data is available for all disciplines, the sample used for estimation does not include disciplines that were never in the National Priority band. Given this data availability, an obvious candidate methodology is difference-in-differences (DID). However, the parallel trends assumption does not seem to hold when comparing the time series of the treatment and candidate control groups (not shown). Different disciplines often display markedly different trends which---in absence of richer and more disaggregated data---will lead to biased estimates. Moreover, the ``no spillover effects'' assumption does not seem to hold either. This is because, by design, the disciplines chosen for the control group would have to be similar to the ones in the treatment group. This very similarity implies that a student might respond to an increase in the price for a degree in the treatment group by choosing another similar degree in the control group. This would violate the ``no spillover effects'' assumption.

Instead of a difference-indifferences approach, I choose one close to an event study. I define the event as ``discipline $i$ is removed from the National Priority band (or, equivalently, it experiences an increase in student contribution)''. There are two instances of this event: one in 2009 and a second in 2012, as detailed above. Furthermore, I restrict the sample to (i) the ever-treated disciplines, i.e. the disciplines that were in the National Priority band at least once; (ii) a time window spanning between 3 years before each event and 5 years after. The size of the window is determined by how many periods are available around both instances of the event. I use this sample to test whether, under an income-contingent loan regime as HECS, drastically increasing the maximum student contribution for a discipline---i.e. the cost to study it---is followed by significant changes in enrolment trends in that discipline. 

\subsection{Empirical specification}
To test this hypothesis, I estimate the following fixed effect equation for discipline $i$ and time $t$:
\begin{equation}
ln(Y_{i,t}) = \alpha_{i} +  \beta_1 T_t + \beta_2 D_{i,t} + \delta D_{i,t} * T_t + \epsilon_{i,t}
\label{fe}
\end{equation}
where $ln(Y_{i,t})$ is the natural logarithm of the number of commencing students by discipline and year, $T_t$ is a yearly event time variable\footnote{An event time variable is a variable equal to the number of periods before/after the event. It takes negative values before the event, positive values after the event and it equals to zero in the year of the event.}, $D_{i,t}$ takes a value of one after the discipline is no longer in the National Priority band and zero otherwise and $D_{i,t} * T_t$ is their interaction term. The coefficient of interest is $\delta$, which can be interpreted as the yearly percentage change in enrolments following a sharp increase in student contribution. 

\section{Results}
\subsection{Graphical analysis}
\begin{figure*}[t]
\begin{subfigure}{.45\linewidth}
  \centering
  \includegraphics[width=\linewidth]{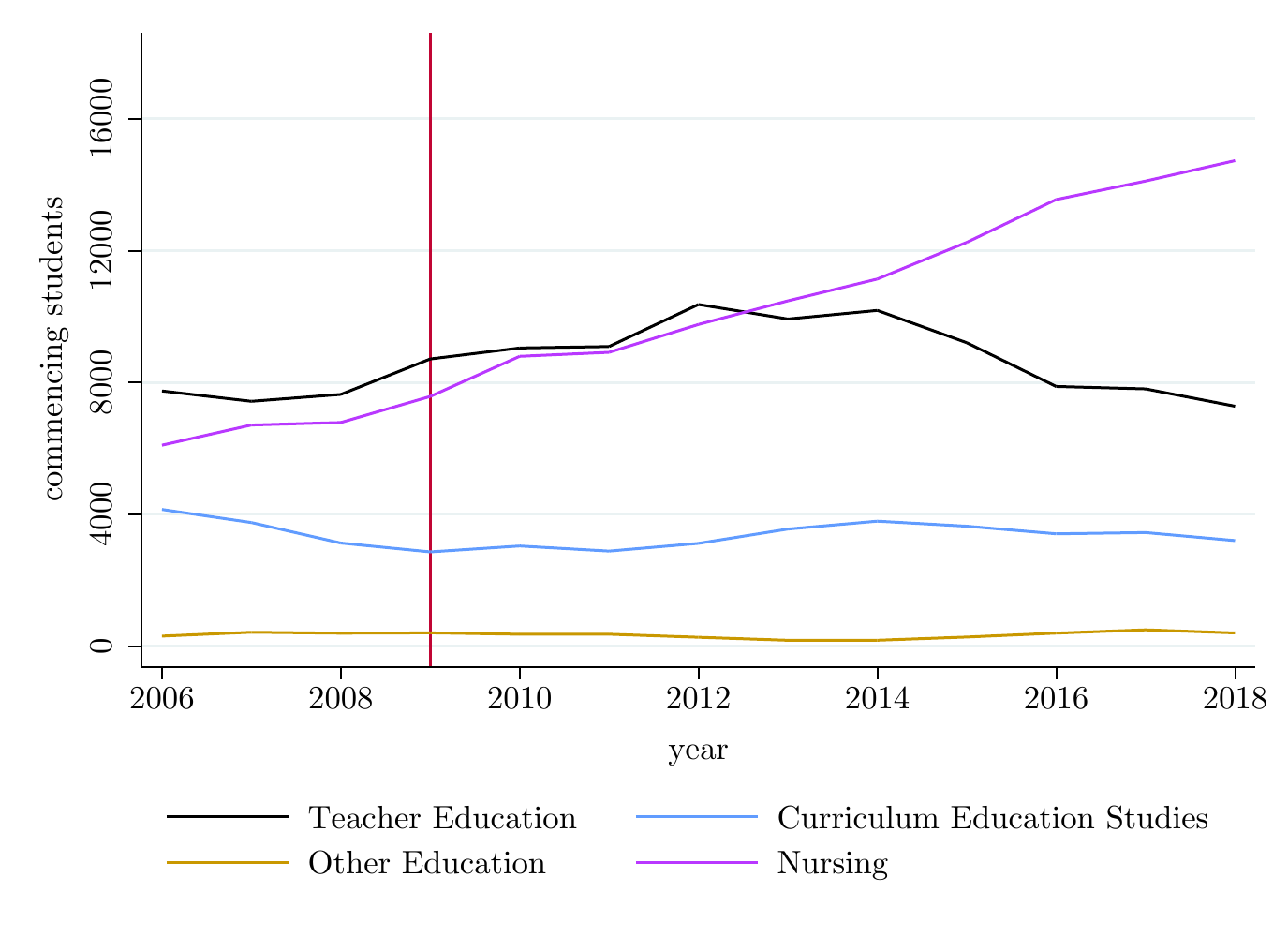}  
  \caption{Nursing and Education}
  \label{fig:nurse}
\end{subfigure}
\hfill
\begin{subfigure}{.45\linewidth}
  \centering
  \includegraphics[width=\linewidth]{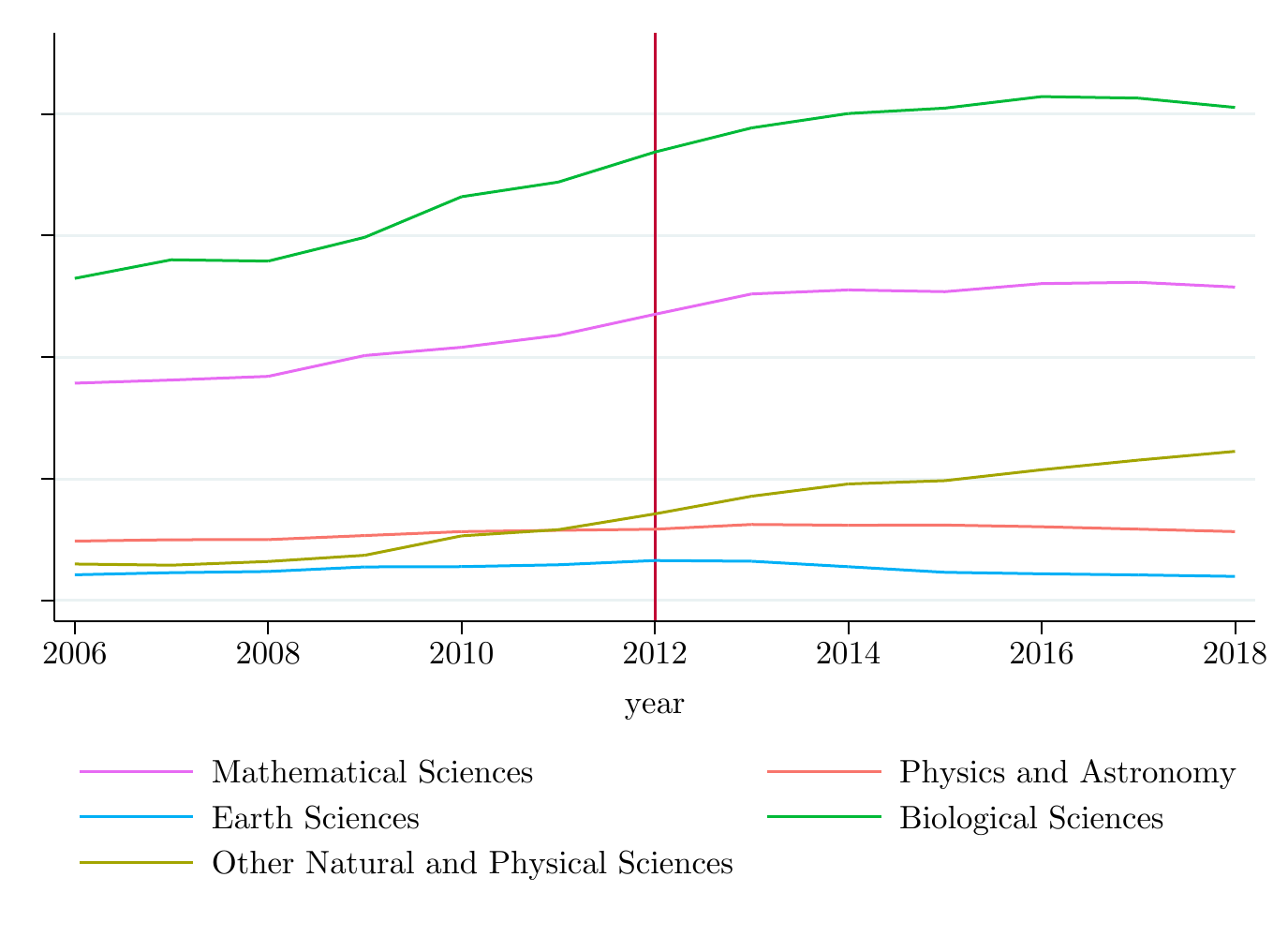}  
  \caption{Physical and Natural Sciences}
  \label{fig:science}
\end{subfigure}
\hfill
\caption{Termination of National Priority}
\label{fig:main}
\end{figure*}

Figure \ref{fig:main} shows the number of commencing students enrolled by disciplines and year. The disciplines shown are those which have been in the National Priority band at least once. Panel (a) shows the number of yearly commencing students who benefited from the National Priority band between 2005 and 2009 \-- where the year 2009 is marked by a red vertical line. The National Priority band between these years included disciplines related to Nursing and Education. Panel (b) shows the number of yearly commencing students who benefited from the National Priority band between 2009 and 2012 \--- 2012 also being marked in red. The National Priority band during these years included all Physical and Natural Sciences, plus Mathematics.

An inspection of Figure \ref{fig:main} suggests that commencing students were rather insensitive to the large changes in student contribution marked by the end of the National Priority programmes. In other words, the demand for those disciplines seems inelastic to changes in price. The only discipline not displaying an uninterrupted linear trend is \textit{Teacher Education}. The number of Teacher Education commencing students peaks in 2012, after which a decreasing trend starts. This change in trend occurs several years after the change in student contribution, hence the two are unlikely to be related.

\subsection{Regression results}
\begin{figure}[b]
\centering
\caption{Event window}
\begin{threeparttable}
 		\includegraphics[width=.9\linewidth]{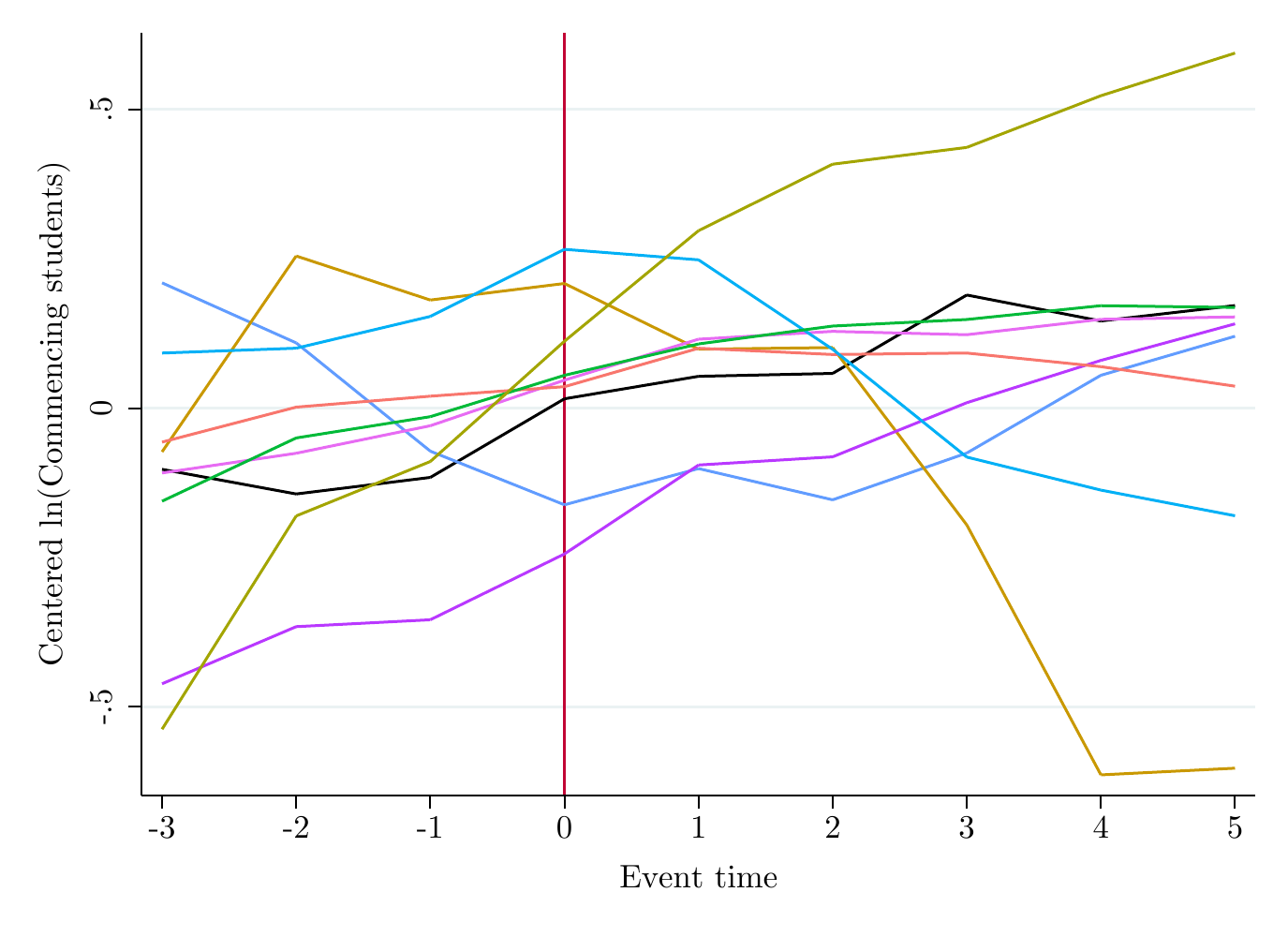}  
		\begin{tablenotes}
			\footnotesize
			\item \textbf{Note.} This figure shows the centered log-transformation of commencing students by year and discipline. These disciplines were  all at least once in the National Priority band. Event time zero is the time at which the event ``being removed from the National Priority band'' occurs. The colour coding is identical to Figure \ref{fig:main}'s, please refer to that legend.
		\end{tablenotes}   
\end{threeparttable}
\label{au_event}
\end{figure}

Column 1 in Table \ref{edu_reg} shows the estimates for Equation \ref{fe}. As anticipated by the graphical analysis, on average there is no significant change in the trend of enrolments after a sharp increase in student contribution ($\delta=-0.06$ with a p-value of 0.16). In Column 2, Equation \ref{edu_reg} is estimated in the restricted case where $\beta_{2}=0$. In other words, in Column 2 the intercept is not allowed to vary after the event. Column 3 and 4 estimate the same equations as columns 1 and 2, respectively, but dropping the field ``Other Education" from the sample. While the coefficient of interest is never significantly different from zero, Columns 3 and 4 show that ``Other Education" is responsible for half of the size of the coefficient of interest, $\delta$. When ``Other Education" is dropped, $\delta$ reduces to less than $0.04$ (in both Columns 3 and 4) and the p-values double to $0.32$ (also in both columns). The leverage of ``Other Education" can also be noticed upon inspecting Figure \ref{au_event}, which visualises the sample after trimming and centering the outcome variable, $ln(Y_{i,t})$. 
\begin{table}[t]
\centering
\begin{threeparttable}
\caption{Changes in enrolments after an increase in student contribution}
\scalebox{.85}{\begin{minipage}{\hsize} \centering
        \input{edu_reg2.tex}		\begin{tablenotes}
			\footnotesize
			\item \textbf{Note.} Column 1 estimates of Equation \ref{fe}. A linear model with discipline fixed effects is fit to data on the logarithm of yearly commencing students by discipline. The specification controls for a linear event time trend, which is interacted with the treatment dummy variable. A discipline is defined as treated after it is no longer in the National Priority band. Column 2 restricts Equation \ref{fe} to the case of no change in the intercept after the event (i.e. $\beta_{2}=0$). Columns 3 and 4 estimate the same equations as column 1 and 2 respectively but excluding ``Other education". This is done in order to show how ``Other education" alone doubles the size of the (negative) coefficient.
		\end{tablenotes}        
\end{minipage}}
\label{edu_reg}
\end{threeparttable}
\end{table}

\begin{table}[t]
\centering
\begin{threeparttable}
\caption{Testing $\delta=0$ using 90\% confidence intervals}
\scalebox{.85}{\begin{minipage}{\hsize} \centering
        \input{edu_90ci.tex}		\begin{tablenotes}
			\footnotesize
			\item \textbf{Note.} I construct and inspect 90\% confidence intervals for $\delta$, using robust standard errors. This is for testing whether $\delta=0$. The model specifications are identical to the ones associated with Table \ref{edu_reg} and are identically ordered. I am again unable to reject the null hypothesis of $\delta=0$.
		\end{tablenotes}    
\end{minipage}}
\label{edu_90ci}
\end{threeparttable}
\end{table}

As an additional test of $\delta=0$ and following \citep{STEIGER04}, I construct and inspect the 90\% confidence interval for $\delta$ (Table \ref{edu_90ci}). The model specifications are identical to the ones associated with Table \ref{edu_reg}. In each specification, zero is included within the 90\% confidence interval. In Columns (3) and (4), whose specifications exclude the outlier discipline, the confidence interval is even more centered around zero.

\section{Discussion} \label{sec:end}
These results are consistent with the hypothesis that income-contingent loans, such as the one featuring in the HECS system, make the demand for a discipline inelastic to price increases. Even when the price for studying a discipline (the student contribution) increases sharply, the trend in new enrolments does not significantly decrease on average. Therefore, HECS seems to be meeting its goal of allowing students to make decisions about their higher education independently its price.

At the same time, these results suggest that if students have access to income-contingent loans, they will not respond to policies that increase the price of a discipline to disincentivise them from choosing it. In light of the evidence provided, the economic argument supporting this kind of policies seems weak, leaving a justification based on moral grounds as the only avenue for a government to support them\footnote{\citep{CHAPMAN97} calls this issue ``the issue of whether or not it is appropriate for the government to charge individuals based on what the expected direct benefits on average from a course are''.}.

\section*{Declaration of Competing Interest}
I have no conflicting interest to declare

\section*{Acknowledgements}
I thank Andrew Norton for his help in navigating the Australian education policy world and for sharing his data, which allowed me to start working on this project quickly. I thank Oliver Maclaren, Daniël Lakens and Guy Prochilo for providing guidance and references on how to provide evidence of absence of an effect rather than an absence of evidence.
\bibliographystyle{apalike}
\bibliography{edu_biblio}

\end{document}

%% file: edu_reg2.tex
{
\def\sym#1{\ifmmode^{#1}\else\(^{#1}\)\fi}
\begin{tabular}{l*{4}{c}}
\toprule
                    &\multicolumn{1}{c}{(1)}         &\multicolumn{1}{c}{(2)}         &\multicolumn{1}{c}{(3)}         &\multicolumn{1}{c}{(4)}         \\
\midrule
T                   &      0.0523         &      0.0632\sym{*}  &      0.0496         &      0.0591         \\
                    &[-0.01,0.11]         & [0.00,0.13]         &[-0.02,0.12]         &[-0.01,0.13]         \\
                    &      (0.03)         &      (0.03)         &      (0.03)         &      (0.03)         \\
\addlinespace
\addlinespace
Event               &      0.0651\sym{**} &                     &      0.0574\sym{**} &                     \\
                    & [0.03,0.10]         &                     & [0.02,0.09]         &                     \\
                    &      (0.02)         &                     &      (0.02)         &                     \\
\addlinespace
\addlinespace
Event{$\times$}T    &     -0.0616         &     -0.0616         &     -0.0363         &     -0.0363         \\
                    &[-0.15,0.03]         &[-0.15,0.03]         &[-0.12,0.04]         &[-0.12,0.04]         \\
                    &      (0.04)         &      (0.04)         &      (0.03)         &      (0.03)         \\
\addlinespace
\addlinespace
Constant            &       8.206\sym{***}&       8.231\sym{***}&       8.446\sym{***}&       8.468\sym{***}\\
                    & [8.10,8.31]         & [8.12,8.34]         & [8.37,8.52]         & [8.39,8.55]         \\
                    &      (0.05)         &      (0.05)         &      (0.03)         &      (0.04)         \\
\midrule
Obs.                &          90         &          90         &          81         &          81         \\
\bottomrule
\multicolumn{5}{l}{\footnotesize 95\% confidence intervals in brackets, robust s.e. in parentheses}\\
\multicolumn{5}{l}{\footnotesize \sym{*} \(p<0.05\), \sym{**} \(p<0.01\), \sym{***} \(p<0.001\)}\\
\end{tabular}
}

%% file: edu_90ci.tex
{
\def\sym#1{\ifmmode^{#1}\else\(^{#1}\)\fi}
\begin{tabular}{l*{4}{c}}
\toprule
                    &\multicolumn{1}{c}{(1)}         &\multicolumn{1}{c}{(2)}         &\multicolumn{1}{c}{(3)}         &\multicolumn{1}{c}{(4)}         \\
\midrule
Event{$\times$}T    &     -0.0616         &     -0.0616         &     -0.0363         &     -0.0363         \\
                    &[-0.14,0.01]         &[-0.13,0.01]         &[-0.10,0.03]         &[-0.10,0.03]         \\
\midrule
Obs.                &          90         &          90         &          81         &          81         \\
\bottomrule
\multicolumn{5}{l}{\footnotesize 90\% confidence intervals in brackets}\\
\multicolumn{5}{l}{\footnotesize \sym{*} \(p<0.05\), \sym{**} \(p<0.01\), \sym{***} \(p<0.001\)}\\
\end{tabular}
}